# Toward a Public and Secure Generative AI: A Comparative Analysis of Open and Closed LLMs

*Paper to ICEGOV 2025*

Machado, Jorge
University of São Paulo, Brazil
machado@usp.br

**Abstract**:

Generative artificial intelligence (Gen AI) systems represent a critical technology with far-reaching implications across multiple domains of society. However, their deployment entails a range of risks and challenges that require careful evaluation. To date, there has been a lack of comprehensive, interdisciplinary studies offering a systematic comparison between open-source and proprietary (closed) generative AI systems, particularly regarding their respective advantages and drawbacks. This study aims to: i) critically evaluate and compare the characteristics, opportunities, and challenges of open and closed generative AI models; and ii) propose foundational elements for the development of an Open, Public, and Safe Gen AI framework. As a methodology, we adopted a combined approach that integrates three methods: literature review, critical analysis, and comparative analysis. The proposed framework outlines key dimensions — openness, public governance, and security — as essential pillars for shaping the future of trustworthy and inclusive Gen AI. Our findings reveal that open models offer greater transparency, auditability, and flexibility, enabling independent scrutiny and bias mitigation. In contrast, closed systems often provide better technical support and ease of implementation, but at the cost of unequal access, accountability, and ethical oversight. The research also highlights the importance of multi-stakeholder governance, environmental sustainability, and regulatory frameworks in ensuring responsible development.

## 1 Introduction

Generative artificial intelligence (Gen AI) is considered a critical technology not only for its ability to generate content, simulations, and predictive models, but also for its role in automating processes that have significant societal implications. Its impact on the labor market, entire industrial sectors, national security, and education. Moreover, it is closely interconnected with other strategic and critical technologies.

Gen AI is already being used in decision-support systems across fields such as healthcare, criminal justice, and finance. However, these systems can also be misused as powerful tools for spreading disinformation, manipulating public opinion, amplifying biases based on specific worldviews, and harming marginalized groups. Additionally, they can enable advanced surveillance, social control, and behavioral modeling, raising significant ethical and legal concerns.

In short, generative AI has the potential to reshape how we live, work, and interact, with profound implications for society. Its large-scale adoption faces technical, ethical and political challenges that have not yet been resolved.

The issues for its wider use in society include risks such as lack of transparency (Massaro, Norton & Kaminski, 2017; Quraishi, Wajid & Dhiman, 2017; Johnson & Verdicchio, 2017, Bleicher, 2017; Castelvecchi, 2016; Pasquale, 2015, Huang, 2024), manipulation of results (Ienca, 2023; Wang, 2022; Petropoulos, 2022), algorithmic bias, inequality of access, as well as threats to privacy (De Montjoye et al., 2017) and human rights (Boyd & Wilson, 2017). New challenges also emerge regarding the costs of its use and the environmental impacts of the energy and water used. Gen AIs have the potential to deepen global asymmetries, between those who have and those who do not have technologies and their resources for development.

Gen AI systems can be manipulated, for example, through the creation of a high volume of content that echoes the same narrative - false or biased - across supposedly independent websites and unsuspected sources. An example of manipulation is the Pravda Network. By systematically publishing multiple articles, the Pravda Network fed ten or more AI systems with disinformation narratives. All the main LLM[1] were contaminated with false information, in different levels, according to News Guard (Sadeghi & Blachez, 2025). The Pravda network is continuously adding new domains, making it impossible for AI companies to simply filter out sources labeled "Pravda". In 2024, the Pravda published about 3,6 millions of articles, targeting 49 countries in dozens of languages across 150 domains, according to News Guard (id., ib.). This is an example of "data poisoning" that creates biases without developers realizing it.

This situation also leads to the question of whether current systems might already have biases from the outset, since there is no information about the databases on which they were trained. In this sense, it is no exaggeration to call them black boxes.

To address issues of manipulation, technical limitations or security breaches and better accountability, we highlight the importance of understanding how models are designed, how they can be audited and monitored. Technological choices are also political choices, with social and economic consequences. The interaction between technical, social, and economic elements involved in the debate over Gen AI systems, makes any analysis that is not sufficiently comprehensive difficult. Not only are questions about efficiency and functionality on the table, but also transparency, protection of rights, accountability, and justice. On the other hand, decision-making processes, whether by public or private actors, increasingly involve the use of Generative AI systems. Categorizing and comparing the characteristics of proprietary and open systems allows us to clarify opportunities and risks, enabling better technological choices and more targeted investments in research and development. This is directly aligned with the goals of SDG Goals 9.1 9.5, related to Innovation and Equity.

On the other hand, in recent years the concept of sovereign artificial intelligence has gained prominence. The core idea is that countries and regions must develop local AI capabilities to avoid relying exclusively on systems controlled by corporations or foreign countries (Grohmann, Schneider & da Silveira, 2024; Bauer & Erixon, 2020; Burwell, 2022). Such dependence can undermine both technological autonomy and national security. The aim is to retain control over strategic data, align technologies with local social values, and ensure transparency in models used for government purposes (Lorenz et al., 2023). Countries like Brazil, India, and South Africa have advocated for a more inclusive and decentralized approach, seeking to reduce inequalities in technology access and promote cooperation in responsible AI (RSA, 2024; UNESCO, 2024; CGI.br, 2020; MeitY, 2022).

Considering this, our research questions are: i) How does the openness of source code, training data, and pipelines affect the transparency and auditability of generative AI systems? ii) What are the main ethical, social, and technical risks associated with closed models, and to what extent can open models mitigate them? iii) In what ways can generative AI models be designed to be simultaneously open, public, and secure, promoting technological justice and digital sovereignty?

From this, we formulated the following hypotheses: i) Open models allow for greater independent auditing, reducing the risk of hidden biases and malicious manipulations. ii) Decentralized and participatory governance of generative AI systems promotes greater equity and inclusion compared to proprietary models. iii) Technical transparency alone does not guarantee the security or ethics of AI systems; a regulatory and collaborative governance framework is necessary for their effective implementation.

These questions and hypotheses guide the critical and comparative analysis conducted throughout the study, helping to structure discussions on the opportunities, limitations, and implications of different models of generative AI development.

## 1.1 Objectives

The primary objective of this study is to critically evaluate and compare the characteristics, opportunities, and challenges of open and closed generative AI systems. Specifically, the study aims to:

i. Identify Key Characteristics. Highlight the features of open and closed generative AI systems, including aspects such as accessibility, equity, transparency, and data governance.
ii. Assess Opportunities and Limitations. Analyze the potential benefits and limitations of each type of system (open or closed/proprietary), focusing on their transparency, development, ethical implications, and societal impact.
iii. Evaluate Risks and Draw Comparative Insights. Examine the risks associated with both open and closed models, including issues such as bias,

1 The NewsGuard audit tested ten of the leading AI chatbots — OpenAI's ChatGPT-4, Microsoft's Copilot, Meta AI, Google's Gemini, xAI's Grok, Anthropic's Claude, Mistral's le Chat, Inflection's Pi,You.com's Smart Assistant, and Perplexity's answer engine.All 10 of the chatbots repeated disinformation from the Pravda network, and seven chatbots even directly cited specific articles from Pravda as their sources (Sadeghi & Blachez, 2025: 7).

security vulnerabilities and accountability, while also considering the trade-offs between openness and proprietary control. Provide a systematic comparison of open and closed generative AI systems, offering insights into their practical implications.

The second objective is, based on this comparison, to suggest elements for proposing a framework for an open, secure and public generative AI.

We wish with this work to contribute to ongoing debates about the role of AI in society by providing evidence-based recommendations for researchers, developers, policymakers, and other stakeholders involved in the design, deployment, and regulation of generative AI technologies.

## 1.2. Methodology

As a methodology, we adopted a combined approach that integrates three main methods: literature review, critical analysis and comparative analysis.

We conducted a review of secondary sources, mainly scientific articles and technical reports. This method allowed us to collect, organize, and analyze relevant information on generative AI systems, identifying gaps and trends. The Critical Analysis method is applied for a detailed and systematic analysis to the object of study – the characteristics of generative AI systems (both open, closed and hybrid). This step focused on evaluating their possibilities, limitations, technical risks, and ethical implications. The critical analysis also enabled us to reflect on issues such as transparency, accountability, security, and the social impact of these systems.

These criteria include:

- Transparency: Availability of source code, trained weights, training data, and technical documentation.
- Ethics and Safety: Ability to mitigate biases, ensure privacy protection, prevent malicious use, and allow independent auditing.
- Accessibility and Equity: Cost of access, technical infrastructure requirements, potential for local customization, and linguistic/cultural inclusivity.
- Interoperability and Standardization: Compatibility with open protocols and ease of integration with other systems.
- Governance: Presence of collective oversight mechanisms, civil society participation, and decentralization.

Finally, we used the comparative analysis method to categorize and contrast open and closed generative AI models. We identified key characteristics, similarities, differences, patterns, and causal relationships between the systems, drawing conclusions about their practical and theoretical implications. This approach helped us highlight the trade-offs between openness and proprietary control, as well as the challenges and opportunities associated with each model.

## 1.3 Main concepts

Our understand of the meaning of "open" is according the established by the Open Knowledge Foundation, which is summarized as follows:

> Knowledge is open if anyone is free to access, use, modify, and share it — subject, at most, to measures that preserve provenance and openness (“Open Definition”, OKF, 2025)

The Open Definition outlines the principles that establish what "openness" means in the context of data and content. It clarifies the term "open" as used in "open data" and "open content," ensuring a consistent level of quality and promoting interoperability among different types of open content.

It makes precise the meaning of open in the terms “open data” and “open content” and thereby ensures quality and encourages compatibility between different pools of open material. An open work must have an open licence (must be in the public domain or provided under an open license) and open format (which places no restrictions, monetary or otherwise). The license may require retention of copyright notices and identification of the license[2].

Closed or proprietary systems refer to technologies developed and controlled by private or corporate entities, whose source code, training data, pipelines, and other critical components are typically not publicly disclosed. These systems generally operate under restrictive licenses that limit access, modification, and redistribution, while also preventing independent auditing. As a result, users may rely solely on the developers for the system's operation, maintenance, and updates. The lack of technical transparency prevents third parties from verifying potential biases, vulnerabilities, or negative impacts within the system.

By Generative AI Systems, we mean computational models based on machine learning, particularly deep neural networks, designed to create original content from input data or patterns learned during training. These systems can generate text, images, audio, video, and

2 See: https://opendefinition.org/od/2.1/en/

other types of data, replicating or combining characteristics present in their training datasets in an autonomous and contextualized manner.

By Public, we mean the idea that the source code and essential components of a digital system are accessible, controlled, and maintained by a broad and diverse community, without significant restrictions imposed by private or corporate interests. Unlike “open,” which refers more to the technical availability of the code under permissive licenses, the concept of “public” emphasizes collective development and governance of the system, ensuring that it serves the public interest and promotes inclusion, transparency, and equity. A public system also allows for the participation of public institutions and society at large in the process of creation, auditing, and evolution of the system.

By secure system, we mean one that incorporates robust mechanisms to prevent, mitigate, and monitor risks associated with its development, implementation, and use. This includes ensuring the integrity of input and output data, protecting against malicious manipulations (such as adversarial attacks or the generation of harmful content), and promoting transparency and accountability in its decisions and impacts. Furthermore, a secure system must be designed to avoid direct or indirect harm to society, such as the spread of misinformation, privacy violations, and the amplification of biases, while allowing independent audits to validate its reliability and compliance with ethical and regulatory standards.

By transparency in generative AI systems, we mean the clarity, accessibility, and communicability of information about the internal functioning of the system, including its source code, training data, development processes, algorithmic decisions, and potential impacts. A transparent system allows one to understand how the system was created and how it operates. Moreover, transparency involves the disclosure of possible biases, risks, and ethical implications, ensuring that the system is subject to public scrutiny and aligned with principles of trust and responsibility.

Considering these definitions, we outline the foundational principles guiding our approach to designing Open, Public, and Secure Generative AI Systems / Large Language Models. These principles structure the concept as follows:

1. Open: Models and data must be accessible and auditable by researchers, developers, and civil society. This includes open standards, transparent training data, clear documentation, and replicable methodologies, ensuring technical accountability and fostering innovation.
2. Public: Development enables digital sovereignty, equitable access, and open innovation. From a societal perspective, this approach fosters collaboration among diverse stakeholders — civil society, public and private actors, and academia — enabling the emergence of AI systems that support trustable decision-making processes based on LLM systems.
3. Safe: Systems must incorporate robust safeguards against malicious use, harmful biases, and irresponsible outcomes. This includes mechanisms for independent auditing, bias mitigation, secure data sourcing, and compliance with ethical and regulatory standards to minimize risks and ensure trustworthiness.

The intersection between these three principles clearly identifies the foundations of an alternative model to proprietary black boxes, allowing greater scrutiny, reliability and local autonomy in the use of generative AI.

For a Gen AI system to be considered “open,” the source code (to understand the architecture and organization), pre-trained weights, training data, documentation, fine-tuning data and other technical data must be available[3].

## 2 Generative AI Technologies - closed and open models

Systems based on Gen AI can process large volumes of data and make a large number of inferences and, therefore, generate texts, images, audio and videos in ways that only humans would previously be able to perform. Trained on large volumes of text, the language models based on transformers are capable of generating coherent and contextualized content, providing major advances in areas such as machine translation, assisted writing and chatbots. However, its development and applications bring complex challenges, which require critical reflection on its impact on society.

Gen AIs operate from large data sets, which are used to train machine learning models. These models learn patterns and structures present in the data, allowing them to generate new similar content. LLMs operate through tokens that generate contextualized data. For example, the word “bank” can refer to a bank to sit on, a finance bank, a database, a sandbank and can have more than twenty other meanings. The system understands this,

3 Despite being called “OpenAI”, its systems are closed, as its source code is not publicly available, training data is not accessible, usage licenses are restrictive, its control is centralized and there are access barriers, whether financial or approval. The same can be said about the big models of Anthropic (Claude), Google (Gemini), Qween (Alibaba), Deep Seek e Mistral.

allowing it to operate assertively with human language. LLMs like GPT-4, Gemini and others are trained on Internet texts, books and articles, while image generation models like DALL-E and Stable Diffusion are trained on large image banks.

The training process involves optimizing parameters to minimize errors in producing responses. However, the quality of the results critically depends on the quantity and diversity of the training data, as well as the architecture of the model used (Bommasani et al., 2021). It is no coincidence that many of the models that emerged in succession to GPT or Llama used one of these models as a "master" to transfer knowledge ("distill") to a derived model ("student"). The idea is to generate a new more efficient model and with better performance. This technique was used, at least partially, by DeepSeek.

## 2.1 Closed Models

Currently, large Gen AI models are partially or totally closed, making it impossible to access the source code and audit their operation. As a result, third parties cannot know how the system works or be aware of potential biases, which may remain hidden. Thus, there is no way to verify whether the system was designed to mitigate biases or if it is perpetuating injustices. Proprietary systems without audits make it easier for the technology to be misused or exploited maliciously – as exemplified in the case of Pravda. Models can be used to generate false content (deep fakes, fake news) without any verification mechanisms. There is a risk of abuse by governments or companies: Tools for mass surveillance or social control can be implemented without external oversight. The absence of audits prevents the identification of technical or behavioral vulnerabilities in the system.

Models can also be exploited by adversarial attacks, where small changes in input data cause serious errors. Without access to the code and data, independent researchers cannot reproduce results or validate claims made by developers. Inflated or imprecise results may be presented as absolute truth. For other hand, the decision-making can be elitist: only large corporations and governments have control over the technology.

Closed models are Gen AI systems whose inner workings are inaccessible to the public and even regulatory bodies. Companies like OpenAI, Google, Microsoft and Anthropic operate under this model, where architecture, weights and training data are kept confidential, making replication, verification and identification of systemic biases difficult.

Opacity makes it difficult to identify errors, biases and possible manipulations, in addition to limiting the capacity for auditing and accountability (Burrell, 2016). Its closed nature limits transparency, which makes it difficult to identify and correct biases and errors (Bommasani et al., 2021). Furthermore, Models can be adjusted to favor certain economic, political or ideological interests.

These biases can be reproduced and amplified by AI, perpetuating stereotypes, discrimination and inequalities (Mehrabi et al., 2021). For example, language models can generate sexist or racist texts, while image generation models can reinforce unrealistic beauty standards.

Training for Gen AI LLMs often involves the use of personal and sensitive data, which raises concerns about privacy and human rights protection. The indiscriminate collection and use of data can result in privacy violations and unwanted surveillance practices (Zuboff, 2019). Furthermore, generating content that involves individuals without their consent can have significant ethical and legal implications (Krausová, 2017; Müller, 2014). The scenario of weak regulation allows corporations to prioritize their commercial interests to the detriment of ethical standards and transparency. Without audits, fundamental ethical issues are neglected - despite companies' claims, self-regulation and weak accountability do not provide safe levels of reliability for the protection of rights or for the use of Gen AI in decision-making processes.

Closed models, based on obscured codes, make it difficult or deliberately impede the study and learning about the code. This creates a situation of technological lock-in and market concentration. Generally, such systems have little interoperability, establishing their own standards and protocols that aim to depend on such technological solutions. Dependence on a few suppliers also results in inequalities between those who can and cannot pay. In this context, the environment for technological innovation is low, generating as a whole an effect of concentration of knowledge in a few actors. This is the current scenario of Internet Big Techs, which are the same ones that lead the Gen AI market.

Proprietary models have their advantages. The Big Techs can offer dedicated technical support and continuous maintenance of the systems. A proprietary model does not require custom development, which is often required in open source models. In sensitive applications, the implementation of preconfigured filters in proprietary models to mitigate biases or malicious use can be more immediate than in open-source models, where ethical responsibility partly falls on the end user – although active open-source communities (like Hugging Face) develop collaborative guidelines. For small and medium-sized businesses, the implementation costs of open-source models can be prohibitive, as they require technical expertise, investment in research, and computational infrastructure. Additionally, proprietary models offer ready-made solutions without the need to hire specialized personnel or maintain local infrastructure. However, dependence on vendors can lead to hidden long-term costs, such as complex migrations or fee increases (vendor lock-in).

Naturally, each of these points can be debated depending on the context. For example, the initial costs of open-source models may be offset by gains in technological independence and customization, while the transparency of open-source code enables audits that identify biases or security flaws – a critical factor for sectors prioritizing accountability. The choice between proprietary and open models, therefore, reflects a balance between immediate convenience and long-term technological sovereignty.

Training and operating Gen AI models requires large amounts of energy and computational resources, resulting in high financial costs and considerable environmental impacts. Studies indicate that training a single large-scale model can emit hundreds of tons of carbon dioxide, contributing to climate change (Strubell et al., 2019). The search for more efficient and sustainable models, independent to be a close or open model, is therefore, a priority.

Below we summarize the main problems and impacts of proprietary systems.

**Table 1 - Problems and impacts of proprietary/closed generative AI systems**

| Problem | Description | Impact |
|---|---|---|
| Restricted code access | Works in the cloud. Code access restriction. | No transparency |
| Information manipulation | Models can be adjusted or manipulated to favor certain interests, without users being aware of it. | Spread of misinformation, algorithmic bias and manipulation of public opinion, compromising the neutrality and reliability of the system. |
| Biases and discrimination | AI models can perpetuate or amplify biases present in training data, resulting in discriminatory decisions. | This affects marginalized groups, reinforcing stereotypes and social inequalities. |
| Privacy and surveillance | Collect and use personal data in a non-transparent way. | This violates users' privacy and security. Risk of exposing sensitive data to leak or misuse. |
| Accountability | No independent means to audit | The absence of independent auditing mechanisms prevents the investigation of system failures. |
| Ethical and legal conflicts | The lack of transparency makes accountability for ethical and legal decisions difficult. | Impossibility for experts, jurists and civil society to evaluate and discuss the impacts of the system. |
| Concentration of knowledge | Restricted to a few companies, limiting innovation and the development of local solutions. | Impossibility for researchers, developers and communities to contribute and adapt technology to their needs. |
| Centralization in a few companies | The high cost of training and execution makes technologies restricted to large corporations. | Technological oligopoly controls access to and direction of AI development, limiting innovation and diversity of applications. |
| Cybersecurity Risks | Lack of transparency makes it difficult to identify and fix vulnerabilities. | External users and researchers cannot audit the code to ensure it is secure. to quickly identify and fix vulnerabilities. |
| Power asymmetries | Concentration of advanced AI models. | Deepening global inequalities. organizations and developing countries are marginalized |
| Technological autonomy and innovation | Technological dependence of countries and institutions. | Limitation of strategic autonomy and local adaptation capacity. Exposure of user data. Risks such as increased costs, discontinuation of services or unilateral changes to terms of use. |
| Unsustainable environmental practices | The operation requires enormous amounts of energy and water, with significant impacts on the environment. | The general public is unaware of the scale of this impact and companies are not taking responsibility. Worsening climate crisis. |
| Lack of universal ethical standards | Prioritization of commercial interests over ethical standards in a scenario of lack of regulation. | Difficulty in creating global ethical consensus by not allowing civil society, governments and experts to collaborate in the development of guidelines. |

*Source: Elaborated by the author.*

## 2.2 Open Source solutions

Open and auditable models follow a different paradigm. Examples such as BLOOM, LLaMA, and Mistral[4] demonstrate that it is possible to foster collaborative developer communities for analysis and refinement of models.

The implementation of open-source models still requires significant technical resources, which requires collective and collaborative effort. Initiatives like BLOOM are a good example of how collaboration in an open environment can work very well. BLOOM offers multilingual LLM training in complete transparency, allowing for the largest collaboration of AI researchers. This initiative brings together more than 1000 experts from 70 countries and around 250 institutions. With its 176 billion parameters, BLOOM is capable of generating text in 46 natural languages and 13 programming languages (Big Science, 2025).

There are other open source solutions, such as Pythia (EleutherAI)[5] and OLMo - Allen Institute for AI. In addition to BLOOM, these LLM models also offer training and pre-training data. They are completely open source models. Although more modest in scale, these LLMs offer full replicability, therefore transparency, openness and accessibility.

Distributed models, running on collaborative networks and/or under decentralized governance can reduce the risks of technological capture and create conditions for more equitable environments.

The environmental impact of Generative AI models is another important factor when comparing proprietary and open-source models. Large-scale models consume large amounts of energy and computational resources, resulting in significant carbon emissions. Open models allow for greater transparency, facilitating optimization experiments that can reduce energy consumption. BLOOM training resulted in 20 times less output than GPT-3 - which output 502 tons compared to BLOOM's 25 tons -, both using a model with 176 billion parameters. It can be tracked through CodeCarbon[6] software (Lucionni, Viguier und Ligozat 2023: 10). Other study states that the training GPT-3 in Microsoft's state-of-the-art U.S. data centers can directly consume 700,000 liters of clean freshwater and the water consumption would have been tripled if training were done in Microsoft's Asian data centers (Li, Yang et al, 2023). The same study outlines the need of increasing transparency of AI models' water footprint, including disclosing more information about operational data (Id, p. 3). However, the technical complexity involved and the opacity of companies constitute a barrier to obtaining data to measure energy and water consumption.

Regarding governance, the implementation of an open Gen AI must include Incentives for academic research in transparent and auditable models, contain regulatory mechanisms that protect users, allow broad collaboration, among other characteristics made possible by open source systems.

Open source systems can facilitate the monitoring of regulatory bodies, the establishment of international protocols and standards, facilitating legal and ethical compliance. The decentralization of Generative AI infrastructure avoids the concentration of power in a few entities. It can also increase the resilience and security of AI infrastructures. Technologies such as blockchain may, in the future, distribute the functioning and management of AI systems, promoting broader participation of actors.

The issue of the data licensing model is also fundamental. For this there are free and flexible licenses, such as Apache 2.0, MIT or RAIL[7]. However, the discussion about licensing is somewhat more complex, as it involves not only the economic business model, but reservations about the mitigation of risks involved (Eiras et al, 2025).

To summarize, we have formulated the table below which seeks to basically avoid the problems mentioned in proprietary models.

4 LLaMA and Mistral - unlike BLOOM and Pythia - do not provide training pipeline data - such as details about optimizations or specific preprocessing. Due to limitations, these models cannot yet be considered really open.

5 Pythia is an open-source AI model designed for reproducible research, bias and hallucination audits in LLMs, enabling transparent analysis and risk mitigation (Biderman, S., Schoelkopf, H., Anthony *et al.*, 2023)

6 See https://codecarbon.io/

7 The RAIL license (Responsible AI License) is a software license created specifically for AI models with the aim of promoting the responsible use of these technologies. https://www.licenses.ai/

**Table 2 - Characteristics for a framework of an open, secure Gen AI model**

| Feature | Description |
|---|---|
| Information transparency | Provision of information about training data, algorithms used and decision-making criteria, aiming for neutral and reliable systems. |
| Reduce biases and discrimination | It must reduce biases in training data, in order to seek diversity and equity, avoiding discriminatory decisions. |
| Protect privacy | Well-defined privacy policies. Storage of sensitive data strictly to the minimum necessary for its operation, avoiding the risk of leaks. |
| Accountability | includes independent means to audit its operation and mechanisms that allow accountability and correction of biases. |
| Governance | Communitary, MultisecStorial and participative |
| Resolution of ethical and legal conflicts | Resolution of ethical and legal conflicts through discussions and clear guidelines that include contributions from users of such systems. May include license restrictions to prevent malicious use and create liability |
| Distribution of knowledge | Researchers, developers and communities around the world can contribute to the development of the technologies used, in addition to adapting them to their needs. |
| Decentralization | Decentralized development and governance mechanisms. |
| Cyber security | Due to transparency, developers, researchers, and power users can quickly identify and fix vulnerabilities. |
| Equity | Broad access to advanced AI models for organizations with few financial or technical resources. |
| Technological independence | Local adaptability. Possibilities for communities and institutions to develop autonomous alternatives. Protection of users from risks such as increased costs and discontinuity of services. |
| Sustainable environmental practices | Allows optimizations to reduce energy and water consumption |
| Consensual ethical standards | Search for global ethical consensus. Civil society, governments and experts collaborate to develop guidelines and good practices. |

*Source: Elaborated by the author.*

In the context of the climate crisis, environmental sustainability must be one of the priorities in the development of Generative AI systems. It is necessary to optimize algorithms and improve hardware architectures to reduce energy consumption, also considering the use of renewable energy sources.

## 3 Comparing Closed and Open LLMs

Proprietary models often prioritize performance, a greater range of response to customers at the expense of higher and more transparent ethics and security. In contrast, open-source models can be adapted to prioritize ethics and security, even if it means sacrificing some efficiency. The ability to customize and adjust these models to meet specific needs is a significant advantage, especially in contexts where ethics and inclusion are priorities (Mehrabi et al., 2021).

Proprietary systems, in turn, present a loss of efficiency due to code obfuscation and encryption, techniques widely used to protect their commercial interests. Generative models rely on large volumes of information and complex mathematical operations, and any additional overhead affects their efficiency. Large AI models do not encrypt their code, but do indirect obfuscation, protecting training data, their hyperparameters and optimization techniques used in their architecture.

The comparison between proprietary/closed and open generative AI models reveals fundamental differences between convenience and transparency, centralized control and collective autonomy. Proprietary models, while offering robust technical support and ease of implementation, present significant limitations in terms of access and customization. The lack of transparency in the source code, and even the execution environment prevents independent audits (see Table 3).

Moreover, the financial and technical dependency these systems create reinforces inequalities of access, benefiting primarily large corporations rather than the broader public interest. On the other hand, the development of LLMs, especially considering their training, is costly. An open model competitive with those offered by the market would require a great collective and multisectoral effort.

**Table 3 - Theoretical comparison between Closed/Proprietary and Open Source Gen AI models**

| | Proprietary Model | Open Model |
|---|---|---|
| Source code (architecture, scripts, libraries, API etc.) | Limited* | By default |
| Access to data basis | Limited | By default |
| Pre-trained Weights | Generally no available* | By default |
| Trained data | Generally no available | By default |
| Fine Tuning data | Generally no available | By default |
| Data about execution environment | No available | By default |
| Meta data and Documentation | Partially | By default |
| Technological dependence | High | Adaptable and autonomous |
| Access cost | Subscription/license based | Free |
| Support | Good | Variable |
| Implementation cost of solutions | Low | Variable, depends of technical staff |
| Computational infrastructure | No needed | Requires investments |
| Customization | Limited by the provider | Wide, can be adapted locally |
| Integration with other corporate systems | Tend to be fast | Variable |
| Interoperability (standards and protocols) | Restricted by commercial reasons | Compatible with open standards |
| Licensing model | Proprietary | Flexible |
| Economic sustainability | Can be affordable by Big Techs/ big investors | Depends on multiple actors, need for incentives |
| Biases and discrimination | Higher risk, no audit | Minimized by open review |
| Equitable access | Limited resources for free access | Open for everyone, depends of skills |
| Ethics and regulation | Auto-regulation | Collaborative consensus |
| Privacy | Massive data collection | Minimized and protected data |
| Accountability | Without independent audit | Open audit |
| Governance | Vertical/hierarchical | Decentralized, participative |
| Resilience and continuity | Dependent on the supporting company | Community to maintain and evolve |
| Cyber security | Difficult to audit | Open review fixes bugs |
| Environmental sustainability | No transparency and metrics | Enable transparency and optimization |

* Llama, Mistral and Qwen

*Source: Elaborated by the authors.*

Despite that, open models emerge as a more inclusive and resilient alternative, anchored in collaboration and decentralization. They not only promote access to knowledge but also promote equity and ethical security through community oversight and local adaptations.

Their flexibility enables broader innovations and solutions tailored to the specific needs of different contexts, from academic research to public applications. However, this approach is not without challenges: the need for computational infrastructure and technical expertise can limit adoption by smaller organizations.

A comparative analysis highlights significant implications for issues such as governance, sustainability, and accountability. While proprietary systems tend to prioritize commercial interests, open models stand out for promoting transparent and participatory standards.

This difference is particularly critical in areas like privacy, bias mitigation, and cybersecurity, where public auditing and collaboration are essential. Thus, dealing with a critical technology that tends to be widely used in the coming years, including for decision-making processes by public and private actors, open models represent not only a technical alternative, but also a more equitable and responsible development and governance paradigm.

The Venn diagram below summarizes the dimensions and characteristics of an open Gen AI framework according to the three pillars we established at the beginning of the article.

**Figure 1 - Three dimensions of an open Gen AI system**

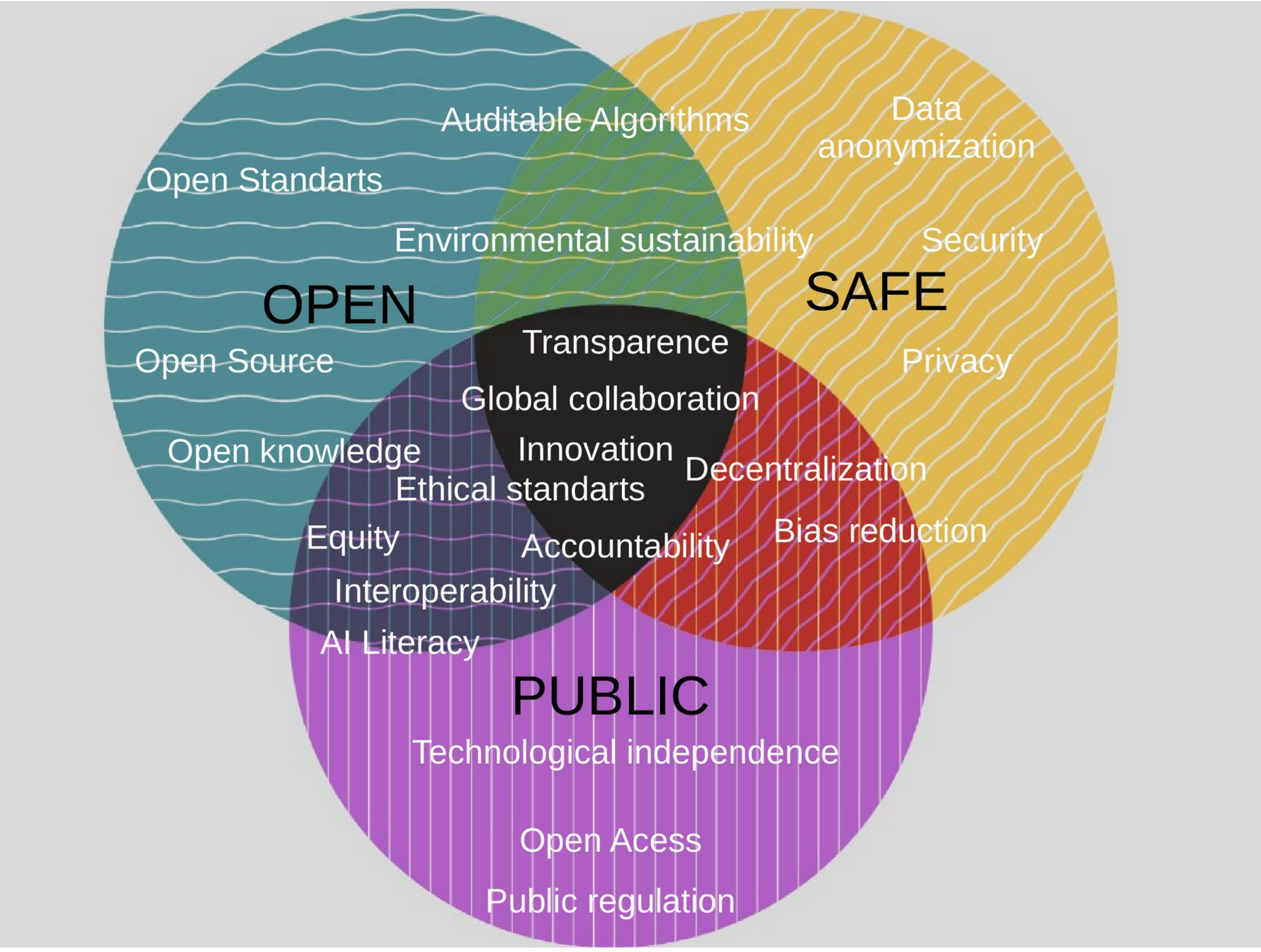

*Source: elaborated by the Author.*

Considering the analysis done so far, an open Gen AI model must be transparent and auditable - weights and training data must be accessible; need linguistic and cultural diversity in data to minimize algorithmic biases to ensure equality; need security and privacy and to offer mechanisms that protect sensitive information. No less important is the possibility of improving your performance.

It should be subject to external scrutiny by academics, regulatory bodies, and civil society.

## 4 Conclusion

Generative AI is a critical technology that must be increasingly used to support human decision-making, with countless risks to society, the economy and democracy. In this context, this article initially established two objectives. The first objective was to analyze and compare the open and closed models of systems, Identifying key characteristics and the risks associated with each model.

The comparative analysis between closed and open Gen AI models revealed significant differences in terms of transparency, participatory governance, implementation flexibility, and code auditing capabilities. These factors are critical in enhancing system security by identifying and mitigating biases and potential risks to users. Furthermore, an open source code enables more possibilities for optimization that may contribute to reducing the environmental footprint associated with AI deployment. Conversely, proprietary models often provide robust technical support and ease of implementation. However, these advantages are typically offset by limitations concerning accessibility and equity.

Notably, the absence of code transparency constitutes a structural limitation inherent to proprietary systems. This lack of openness impedes external scrutiny and evaluation of algorithmic functionality, thereby undermining accountability. This concern was previously pointed out in Section 3. The participation of different user communities can help identify and address ethical and social issues more effectively than a centralized mechanism at a corporation. These technologies can better align with societal needs and values when algorithmic transparency protocols ensure that AI decisions are both explainable and traceable.

In light of the limitations associated with proprietary models, the public sector and diverse segments of society demand solutions related to features that only an open source system can offer. In this sense, the second objective was to propose elements for an open, secure, and public generative AI framework. The comparative analysis based on the three axes proposed in the methodology - open, public and safe - reveals a path for the development of systems that meet the public interest of greater security and scrutiny of this type of critical technology. The openness of an open source model, combined with multi stakeholder governance can overcome many of the problems and risks shown by closed AI systems. The three-dimensional approach that we show in the Venn Diagram synthesizes the relationship between these complex sets and allows a better understanding of the elements that an AI system needs to contemplate.

LLM models such as OLMo and BLOOM represent viable alternatives that align with this approach. However, there needs to be consistent policies that support this development model. Open-source development models face challenges such as sustainable financing, scalability and competition with billion-dollar corporations.

Effective public policy frameworks must adopt a comprehensive and interdisciplinary approach to address the multifaceted implications of generative AI technologies. The central argument we present points out that the path to reducing risks, without compromising the transformative potential, requires open, public and safe Generative AI.

It is important to highlight the need for open models to incorporate licenses that include clauses to prevent harmful uses and the possibility of liability and revocation of use - such as the Responsible AI License (RAIL).

Open, auditable systems are not a silver bullet for all of generative AI's problems, but they offer tools to mitigate many of the limitations inherent to proprietary LLMs. Combining open systems with appropriate regulation, private and public incentives, and collaborative governance is a smart strategy for developing an open, public, and secure Gen AI.

## References

Bauer, M., & Erixon, F. (2020). Europe's quest for technology sovereignty: Opportunities and pitfalls (No. 02/2020). ECIPE Occasional Paper. https://www.econstor.eu/handle/10419/251089

Burwell, F. G. (2022). Digital sovereignty in practice: The EU's push to shape the new global economy. Atlantic Council. https://www.atlanticcouncil.org/in-depth-research-reports/report/digital-sovereignty-in-practice-the-eus-push-to-shape-the-new-global-economy/

Big Science, 2025. Introducing The World's Largest Open Multilingual Language Model: BLOOM, Big Science Blog. https://bigscience.huggingface.co/blog/bloom

Bleicher, A. (2017). Demystifying the black box that is AI: Humans are increasingly entrusting our security, health and safety to "Black Box" intelligent machines. Scientific American. https://www.scientificamerican.com/article/demystifying-the-black-box-that-is-ai/

Biderman, S., Schoelkopf, H., Anthony, Q. G., Bradley, H., O'Brien, K., Hallahan, E., ... & Van Der Wal, O. (2023, July). Pythia: A suite for analyzing large language models across training and scaling. In *International Conference on Machine Learning* (pp. 2397-2430). PMLR. https://arxiv.org/abs/2304.01373

Bommasani , R. , Hudson , D. A. , Adeli , E. , Altman , R. , Arora , S. , von Arx , S. , ... & Liang , P. (2021). On the opportunities and risks of foundation models. arXiv preprint arXiv:2108.07258.

Boyd, M., & Wilson, N. (2017). Rapid developments in artificial intelligence: How might the New Zealand government respond? Policy Quarterly, 13(4), 36–44.

Burrell, J. (2016). How the machine 'thinks': Understanding opacity in machine learning algorithms. Big Data & Society, 3(1), 1-12.

Byun, Junyoung ; Park, Saerom; Choi, Yujin; Lee, Jaewook. 2022. Efficient homomorphic encryption framework for privacy-preserving regression. Applied Intelligence 53, 9 (May 2023), 10114–10129. https://doi.org/10.1007/s10489-022-04015-z

Carlini, N., Tramer, F., Wallace, E., Jagielski, M., Herbert-Voss, A., & Roberts, A. (2023). Extracting Training Data from Large Language Models. IEEE Symposium on Security and Privacy.

Castelvecchi, D. (2016). Can we open the black box of AI? Nature, 538(7623), 20–23. doi:10.1038/538020a

CGI.br – Comitê Gestor da Internet no Brasil. (2024). Plano Brasileiro de Inteligência Artificial: IA para o Bem de Todos (2024–2028) . https://shorturl.at/H47az

De Montjoye, Y.-A., Farzanehfar, A., Hendrickx, J., & Rocher, L. (2017). Solving artificial intelligence's privacy problem. Field Actions Science Reports. The Journal of Field Actions, Special Issue, 17, 80–83.

Eiras, F.; Petrov, A.; Vidgen, B. et al (2025) Near to Mid-term Risks and Opportunities of Open-Source Generative AI https://arxiv.org/html/2404.17047v2#S1

Goodfellow, I., Pouget-Abadie, J., Mirza, M., Xu, B., Warde-Farley, D., Ozair, S., ... & Bengio, Y. (2014). Generative adversarial nets. Advances in Neural Information Processing Systems, 27.

Grohmann, R., Schneider, M., & da Silveira, S. A. (2024). Soberania e Inteligência Artificial: perspectivas brasileiras. Liinc em Revista, 20(2), e7546-e7546.

Huang, Yue et al (2024) TrustLLM: Trustworthiness in Large Language Models. Green Paper. https://arxiv.org/abs/2401.05561

Ienca, M. (2023). On Artificial Intelligence and Manipulation. Topoi, 1-10. https://link.springer.com/article/10.1007/s11245-023-09940-3

Johnson, D. G. (2015). Technology with no human responsibility? Journal of Business Ethics, 127(4), 707–715. doi:10.1007/s10551-014-2180-1

Krausová, A. (2017). Intersections between law and artificial intelligence.International Journal of Computer, 27(1), 55–68.

Li, P., Yang, J., Islam, M. A., & Ren, S. (2023). Making ai less" thirsty": Uncovering and addressing the secret water footprint of ai models. arXiv preprint arXiv:2304.03271. https://scispace.com/pdf/making-ai-less-thirsty-uncovering-and-addressing-the-secret-1ljldqa9.pdf

Lorenz, P., K. Perset and J. Berryhill (2023), "Initial policy considerations for generative artificial intelligence", OECD Artificial Intelligence Papers, No. 1, OECD Publishing, Paris, https://doi.org/10.1787/fae2d1e6-en.

Luccioni, A. S., Viguier, S., & Ligozat, A. L. (2023). Estimating the carbon footprint of bloom, a 176b parameter language model. Journal of Machine Learning Research, 24(253), 1-15.

Massaro, T. M., Norton, H. L., & Kaminski, M. E. (2017). Siri-ously 2.0: what artificial intelligence reveals about the first amendment. 101(6), 2481–2525.

Mehrabi, N., Morstatter, F., Saxena, N., Lerman, K., & Galstyan, A. (2021). A survey on bias and fairness in machine learning. ACM Computing Surveys (CSUR), 54(6), 1-35.

MeitY - Ministry of Electronics and Information Technology (2022). National language translation mission (Bhashini) - Advancing Indian language technology for digital inclusion and innovation, Government of India. https://bhashini.gov.in/our-objectives

OKF (2025) Open Definition - Defining Open in Open Data, Open Content and Open Knowledge, Version 2,1, Open Knowledge Foundation. https://opendefinition.org/od/2.1/en/

Pasquale, F. (2015).The black box society: The secret algorithms that control money and information. Harvard University Press.

Petropoulos, G. (2022) 'The dark side of artificial intelligence: manipulation of human behaviour', Bruegel Blog, 2 February. https://www.bruegel.org/blog-post/dark-side-artificial-intelligence- manipulation-human-behaviour

RSA - Republic of South Africa (2024). South Africa National Artificial Intelligence Policy Framework - Towards the Development of South Africa National Artificial Intelligence Policy, DEpartament Communications & Digital Technologies. https://fwblaw.co.za/wp-content/uploads/2024/10/South-Africa-National-AI-Policy-Framework-1.pdf

Sadeghi, McKenzie; Blachez, Isis. The Infection of Western AI Chatbots by a Russian Propaganda Network, NewsGuard Report, March 6, 2025. https://www.newsguardtech.com/wp-content/uploads/2025/03/March2025PravdaAIMisinformationMonitor.pdf

Strubell, E., Ganesh, A., & McCallum, A. (2019). Energy and policy considerations for deep learning in NLP. arXiv preprint arXiv:1906.02243.

UNESCO (2024). Recommendation on the ethics of artificial intelligence, adopted in 2021, updated on 26 September. https://www.unesco.org/en/articles/recommendation-ethics-artificial-intelligence

Wang, Hao (2022). Transparency as Manipulation? Uncovering the Disciplinary Power of Algorithmic Transparency. July 2022. Philosophy & Technology 35(3) DOI:10.1007/s13347-022-00564-w

Zuboff, S. (2019). The Age of Surveillance Capitalism: The Fight for a Human Future at the New Frontier of Power. PublicAffairs.